\documentclass[useAMS,usenatbib,asm]{mn2e}

\usepackage{graphicx,fleqn,times,color}
\usepackage{natbib}
\usepackage{amsmath}
\usepackage{float}

\title[A novel method to bracket the corotation radius in galaxy disks: vertex deviation maps]{A novel method to bracket the 
corotation radius in galaxy disks: vertex deviation maps}
\author[Roca-F\`abrega et al.]{Santi Roca-F\`abrega $^{1}$, Teresa Antoja $^{2,3}$, Francesca Figueras $^{1}$, Octavio Valenzuela $^{4}$,
\newauthor Merc\`e Romero-G\'omez $^{1}$ and B\'arbara Pichardo $^{4}$\\ \\
$^{1}$ Departament d'Astronomia i Meteorologia and IEEC-UB, Institut de Ci\`encies del Cosmos de la Universitat de Barcelona,\\
     Mart\'i i Franqu\`es, 1, E-08028 Barcelona.\\
$^{2}$ Kapteyn Astronomical Institute, University of Groningen, PO Box 800, 9700 AV, Groningen, The Netherlands. \\
$^{3}$ Research and Scientific Support Office, European Space Agency 
(ESA-ESTEC), PO Box 299, 2200 AG Noordwijk, The Netherlands. \\
$^{4}$ Instituto de Astronom\'ia, Universidad Nacional Aut\'onoma de M\'exico, A.P. 70-264,
   04510, M\'exico, D.F.; Ciudad Universitaria, D.F., M\'exico.\\
}

\begin{document}

\date{Accepted 2014 March 4.  Received 2014 March 3; in original form 2014 January 16}
\pagerange{\pageref{firstpage}--\pageref{lastpage}} \pubyear{2014}
\maketitle 

\label{firstpage}

\begin{abstract}
We map the kinematics of stars in simulated galaxy disks with spiral arms using the velocity ellipsoid vertex deviation (l$_v$).
We use test particle simulations, and for the first time, fully self-consistent high resolution N-body models.
We compare our maps with the Tight Winding Approximation model analytical predictions.
We see that for all barred models spiral arms rotate closely to a rigid body manner and  the vertex deviation
values correlate with the density peaks position bounded by overdense and underdense regions.
In such cases, vertex deviation sign changes from negative to positive when crossing the spiral arms in the direction of disk rotation, 
in regions where the spiral arms are in between corotation (CR) and the Outer Lindblad Resonance (OLR). By contrast, when the arm sections are inside the CR and outside the OLR,
 l$_v$ changes from negative to positive.We propose that measurements of the vertex deviations pattern can be used to trace the position of the main resonances of the spiral arms.
We propose that this technique might exploit  future data from Gaia and APOGEE surveys.  For unbarred N-body simulations with spiral arms corotating with disk material at all radii,
 our analysis suggests that no clear correlation exists between l$_v$  and density structures.
\end{abstract}


\section{Introduction}\label{sec:intro}

It is well known both from simulations and observations that the large scale structures in galaxies such as bars and spiral arms strongly
 affect the stellar kinematics of their disks. One of the properties of such structures that is most relevant for the disk
 dynamics is their pattern speed which sets the position of the resonance radius. \citet{Rautiainen08} have exhaustively reviewed the
 methods usually
 applied to external galaxies to derive this information. Some of these methods are model independent techniques such as the so-called
 \citet{Tremaine84} method, whereas others are parametric
 - they fit several given analytical potential components to observations \citep[i.e.][]{Zhang07} - or based on the 
relation of various morphological or photometric features with resonances \citep[e.g.][]{Martinez09}. The model dependent methods
 are based on several assumptions, being one of the most critical the adoption of a model for the spiral arm kinematics
 (the density wave theory is usually imposed). In addition to these methods, and thanks to the new detectors available, kinematic methods 
based on the analysis of the residual pattern in the velocity field, have been used by \citet{Canzian93} and recently by \citet{Font11}.
 With these methods, a residual velocity map allows the exploration of 
the resonant structure of the galactic disks but, again, they require the removal of circular velocities (previous knowledge of
 the galactic rotation curve) and, up to now they have been applied only to the gas component.\\

Focusing in our Milky Way galactic disk, the corotation
 radius (CR) of the spiral pattern is still a controversial parameter. Without intending to be exhaustive, CR
 has been estimated, for example, by applying techniques such as the Ogorodnikov-Milne Model to the Local Stellar System Kinematics
 of young Hipparcos stars \citep{Fernandez01} or by evaluating the change of the kinematic substructures of the velocity 
field (usually named as moving groups) in test particle simulations \citep[e.g.][]{Chakrabarty07,Antoja11}.
 Related to this, the distribution of the fine structure in the velocity space has been used to study bars
 and spiral arms morphology and dynamics both from solar neighbourhood observational data \citep[e.g.][]{Dehnen00} and from
more extended RAVE data \citep{Antoja13}. All these methods require, again, the adoption of a 
model for the bar or spiral arm potential.\\
At the eve of the Gaia era, new methodologies have to be set up for such analysis in our Galaxy. Our aim here is 
to analyze deeply the connection between resonant radius and the moments of the stellar velocity distribution
 function. This link will open up new avenues for a model independent kinematic method to determine CR. In 
this context, \citet{Vorobyov06,Vorobyov08} demonstrated, from simulations, how second order moments of the stellar velocity
 distribution can be potential tracers of large scale structures such as the spiral arms in galactic disks. As known, the velocity
 ellipsoid in axisymmetric systems is oriented with the radial and azimuthal axis of the galaxy. The misalignment of this ellipsoid,
 known as vertex deviation (l$_v$), provides important information on the non-axisymmetric components. Following that pioneer
 work, in this first paper we will focus on the analysis of the l$_v$ all through the galactic disk, postponing for further investigations
 the use of higher order moments or even the reanalysis of first order moments, that is the mean residual velocities.\\

The misalignment of the velocity ellipsoid in the solar neighbourhood was initially reported by \citet{Stromberg46}. 
The determinations of the local l$_v$ from Hipparcos data are around 30 deg for early-type stars and 10 deg for old-disk
stars \citep{Dehnen98}. More recent measurements give values around 20 deg, using late type stars from SDSS \citep{Fuchs09}.
 Several possible causes for this non null l$_v$ have been proposed, most of them related, as mentioned, to the non-axisymmetric 
components (bar and spirals). 
These non-axisymmetric components can create patterns in the velocity field that change the shape and orientation of the
 velocity ellipsoid. Furthermore the existence of fine kinematic substructure (moving groups) may also change the moments
of the velocity distribution function \citep{Binney08}. This kinematic substructure can consist of groups of young stars still following
 similar orbits from the time when they were born or can be caused by the resonances of the non-axisymmetries of the Galaxy for example
through mechanisms such as resonant trapping or scattering \citep[e.g.]{Antoja09}.\\

The possible connection of l$_v$ with the spiral structure was early discussed by \citet{Woolley70} and \citet{Mayor70}. The former concluded
 that l$_v$ is a remnant of the conditions of stars when they were formed, mostly based on the fact that l$_v$ is observed in young stars but
 not so clearly in the old populations. The latter used the analytical expressions of the density wave theory from \citet{Lin69}
 to quantify this effect in the solar neighbourhood. He derived analytically the second order moments of the velocity distribution function
by considering a Galactic system composed of an axisymmetric part and a spiral perturbation and using the zero and first
order moments computed by \citet{Lin69}. In particular, the Tight Winding Approximation
 (TWA hereafter) was imposed. As described by \citet{Binney08}, the TWA spiral arms model is the result of using the WKB approximation
 used in quantum mechanics. In \citet{Mayor70}, the epicyclic approximation was adopted and velocity dispersions were assumed to be small 
(valid for a young and cool population). Furthermore, his analytical approach required a small amplitude and pitch angle of the spiral pattern
 (i.e. the TWA shall be fulfilled). Later on, \citet{Hilton82} reproduced the observed l$_v$ sign and magnitude for young stars with a model
 where dense molecular clouds are launched from spiral arms at post-shock velocities, and as a consequence, the forming stars move at the same
 velocities. Coming back to the analytical approach, \citet{Kuijken94} found how elliptical potentials could also lead to a 
non-vanishing l$_v$ and tested their results using orbital integrations. \citet{Muhlbauer03} and \citet{Monari13} showed that a barred
potential induces different l$_v$, depending on the position with respect to the bar and its main resonances. They also found that
the l$_v$ increases with decreasing velocity dispersion.\\
 As discussed above, \citet{Vorobyov06,Vorobyov08} computed the
 moments of the velocity distribution function across the disk in both test particles simulations, imposing a spiral arms’ potential, and in spiral
 arms’ semi-analytical models (with the so called BEADS-2D code). In this study they found a clear correlation between the position of the density 
structures (i.e. the spiral arms) and the change of the sign of the l$_v$. They saw that large regions with positive l$_v$
 are present in front of the spiral arm (following the spiral rotation), while negative l$_v$ were found behind the arms. In their analysis,
 however, only cases where the spiral structure is located outside CR were considered.\\
Here we map the l$_v$ caused by the spiral arms in the whole galactic disk. To undertake our study we use: 
i) an analytical approach, ii) test particle simulations imposing a fixed 2-armed galactic potential or a bar; and, iii) 
self-consistent N-body simulations.
First, in our analytical development, we extend the modelling  of \citet{Mayor70} to analyse the expression for the l$_v$ in the TWA
 model not only locally, as he did, but across the whole disk. Second, the use of test particle simulations allows us to 
control the parameters of the imposed potential and to explore the parameter space and its influence on the results. For example,
 we can fix the position of the spiral pattern resonances at our convenience and monitor the behaviour of the l$_v$ inside
 or beyond CR, which was not addressed in previous studies. Besides, we also use simulations where the spiral arms are formed as a response
 to an imposed barred potential. Finally, N-body simulations,
 used here for the first time to map the l$_v$, provide a more realistic framework because they are fully self-consistent. 
These simulations also allow us to analyse the evolution in time of l$_v$.

In Sect.~\ref{sec:lv} we give the definition of the l$_v$ and the expressions to compute its error. In Sect.~\ref{sec:methods} we present our analytical approach and
 the simulations analysed here for both test particle
and N-body simulations. The results from our analysis are presented in Sect~\ref{sec:lvres} and in Sect~\ref{sec:conc} we
 summarize and give our conclusions. Finally, in Appendix \ref{sec:appendix1} we detail the calculations for the analytical 
expression of the l$_v$ in an axisymmetric potential plus TWA spiral arms.

\section{Vertex deviation}\label{sec:lv}

The velocity dispersion tensor that defines the $(p,q,r)$th centered moments of the velocity distribution at position ${\bf x}$ and time $t$ is defined as:
\begin{eqnarray}
\label{eq:mom}
\mu_{pqr} =  \frac{1}{\mu_{000}}\int d^3 {\bf v} \left(u-\bar{u}\right)^p\left(v-\bar{v}\right)^q\left(w-\bar{w}\right)^r f,
\end{eqnarray}
with ${\bf v}=(u,v,w)$ where $u$, $v$ and $w$ denote, respectively, the radial, azimuthal and vertical velocity components and $f=f({\bf x},{\bf v},t)$ is
the velocity distribution function.
The vertex deviation (l$_v$) is the angle that mesures the tilt of the velocity ellipsoid, in the u-v plane, compared to the 
orientation of an axisymmetric configuration; it is related to a non null value of the cross correlation coeficient $\mu_{110}$. Here we use
 the extended definition presented in \citet{Vorobyov06} that includes the 
possibility of having large l$_v$, which happens when breaking the epicyclic approximation locally in regions where the spiral gravitational potential is strong:
\begin{eqnarray}
\label{eq:lv1}
\tilde{l}_{v} & = & \frac{1}{2}\mbox{atan}\left(\frac{2\mu_{110}}{\mu_{200}-\mu_{020}}\right)
\end{eqnarray}
\begin{eqnarray}
\label{eq:lv2}
l_v & = & \left\{ \begin{array}{ll}
                 \tilde{l}_v & \mbox{if $\mu_{200}>\mu_{020}$} \\
                 \tilde{l}_v + \mbox{sign}\left(\mu_{110}\right)\frac{\pi}{2} & \mbox{if $\mu_{200}<\mu_{020}$}
                      \end{array} \right.
\end{eqnarray}

We computed the error on l$_v$ (denoted by $\epsilon_{l_v}$) as the propagation of the errors in the second and fourth order moments \citep{Nunez82}:
\begin{eqnarray}
\label{eq:lverr}
 \epsilon_{l_v}&=&\left|a_4\sqrt{b_1+a_1b_2+b_3\left[\frac{\mu_{220}}{N}+\mu_{110}^2a_2+\mu_{200}\mu_{020}a_3\right]}\right| \nonumber \\
\nonumber \\
 a_1&=&2\left(N-1\right)^{-1}-3N^{-1} \nonumber \\
 a_2&=&\left(N-1\right)^{-1}-2N^{-1} \nonumber \\
 a_3&=&\left(N-1\right)^{-1}-N^{-1} \nonumber \\
 a_4&=&\left(\mu_{110}\left(a_1+4\right)\right)^{-1}  \\
 b_1&=&\left(\mu_{400}+\mu_{040}\right)N^{-1} \nonumber \\
 b_2&=&\left(\mu_{200}^2+\mu_{020}^2\right) \nonumber \\
 b_3&=&\left(\mu_{200}-\mu_{020}\right)^2\left(\mu_{110}\right)^{-2} \nonumber
\end{eqnarray}
This expression takes into account the fact that the error is larger both, when the number of particles is low - due to Poisson noise -
 and when the velocity ellipsoid is nearly circular so the major axis of the velocity ellipsoid is not well defined.

\section{Methodology}\label{sec:methods}

Here we present our analytical development as well as the characteristics of the simulations used in our analysis. 

\subsection{TWA analytical approach}\label{subsec:analytic}

We have derived the analytical expression for the vertex deviation l$_v(r,\theta)$ of the velocity distribution function proposed by
\citet{Lin69}. This consists of a perturbed classical Schwarzschild distribution, where the perturbation is the result of a m-armed 
\citet{LinShu} spiral arm. The final
expressions and the development procedure are presented in Appendix \ref{sec:appendix1} and they are a generalization of the 
expressions by \citet{Mayor70}. We use these expressions
to map the l$_v$ values across the whole galactic disk.\\
As input parameters we used a spiral arms' rotation frequency of 35 km~s$^{-1}$~kpc$^{-1}$, a pitch angle
 of 8 deg, a mass of 5$\%$ of the disk mass, a radial velocity dispersion of 20 km s$^{-1}$, constant with radius, and a disk rotation curve 
derived from the axisymmetric Galactic model of \citet{Allen91}. The CR in this model is placed at 6.2 kpc
and the OLR at 10.2 kpc, while the ILR does not exist. The amplitude of the spiral arms potential declines in radius as $\propto$~r~$\exp(-\text{r}/\text{R}_{\Sigma})$
with a radial scale length of R$_{\Sigma}$=2.5 kpc. We use an amplitude normalization (A$_{sp}$) of 850 km$^2$~s$^{-2}$~kpc$^{-1}$.
 The locus of the spiral is an m=2 logarithm that starts at 2.6 kpc.

\subsection{Test particle simulations}\label{subsec:test}

We run test particle simulations using several galactic potentials. Potentials used here are the result of a superposition of an axisymmetric
 part plus spiral arms or bar components. In all cases, the axisymmetric
 component is the one described in \citet{Allen91} and consists of the superposition of analytical and time independent bulge, disk and halo potentials.
 Here we analyze separately the non-axisymmetric components of the potential (imposed bar and
 imposed spirals) to avoid a more complex scenario when interpreting the connection of density structures with the l$_v$.

The parameters of our basic models and characteristics of our simulations are presented in Table~\ref{tab:simulations},
 where we show for each model some of the main properties of the potential, the number
 of particles and the total integration time.  The number of particles
 in all cases is around or much higher than 5$\cdot$10$^6$ and the integration time was from about 5 to 20 rotations of the 
non-axisymmetric structure. Whereas with large integration times the test particles have reached approximately total statistical equilibrium
 with the galactic potential, for shorter times the particles may not be completely relaxed or face-mixed. Nonetheless, by analysing snapshots
 with higher and lower evolution times we tested that results presented in Sect.~\ref{sec:lvres} are independent of the integration time.

\subsubsection*{Spiral arm potentials} 

We use two different types of spiral arm potentials, 
namely the TWA \citep{Lin69,Binney08} and the PERLAS \citep{Pichardo03} models, which are both described in \citet{Antoja11}. 
These simulations are for 2D disks.
As initial conditions we used an axisymmetric Miyamoto-Nagai disk density profile. The initial velocity field has been
 approximated using the moments of the collisionless Boltzmann equation, simplified by the epicyclic approximation with a local
 normalisation of $\sigma_u$(R$_{\odot}$)=20 kms$^{-1}$ (for more details see \citealt{Antoja11}). According to \citet{Aumer09} this value 
corresponds to a young population of late B, early A type stars. We refer to these initial conditions as ICMN20.
In some cases we use simulations generated for other purposes \citep[see][]{Antoja11}. They use an initial 2D exponential 
density distribution as in \citet{Hernquist93}, and are named IC2.

The simulations named TWA1, TWA2 and TWA3 are our basic models. The non axisymmetric component of the potential have been introduced
 abruptly from the beginning. Particles have been then integrated during 5 spiral arm rotations.
We also produced experiments increasing adiabatically the spiral arms and we noticed that for the pitch angle and spiral arms mass
 ratio employed here, the effect of introducing the arms impulsively (from t=0) or adiabatically, is negligible.
 The parameters that fix the TWA potential are the amplitude of the cosine perturbation $A(R)=-A_{sp}R\exp(-R/R_{\Sigma})$,
 the radial scale length $R_{\Sigma}$,
 the number of spiral arms $m$, its initial phase $\phi_0$, its locus $g(r)$, which in turn depends on the pitch angle $i$, 
and the radius at which
 the spiral arms begin $R_{sp}$ \citep[see][]{Antoja11}. Here we used $A_{sp}$=850 km$^2$~s$^{-2}$~kpc$^{-1}$, $R_{\Sigma}$=2.5 kpc, 
$m$=2, $\phi_0$=0, $R_{sp}$=2.5 kpc, and
 $g(r)=-2/\tan(i)\ln(r/R_{sp})+\phi_0$. The pattern 
speed for TWA1 (TWA2 and TWA3, respectively) is fixed to 20, (35 and 50, respectively) km~s$^{-1}$~kpc$^{-1}$ and it is assumed to 
be constant at all radii. As known a
 change on this parameter directly produces a change in the CR. This lets us study differences on the l$_v$ values
 inside or outside CR. Note that TWA2 has very similar parameters to the ones set in our analytical approach 
of Sect.~\ref{subsec:analytic}.

Apart from the main models (TWA1, TWA2 and TWA3) we performed several more simulations only changing one of the parameter each time.
 This is to test the independence of our 
results on the most critical parameters: spirals amplitude, pitch angle, initial velocity dispersions of the 
test particles, angular speed of the pattern and total integration time.

In particular, we scanned the values of 8 to 15 deg for the pitch
angle, 600 to 1300 km$^2$s$^{-2}$kpc$^{-1}$ for the TWA amplitude, 10 to 40 kms$^{-1}$kpc$^{-1}$ for the radial velocity dispersions and
 2 to 7 spiral rotations for the integration time.

The PERLAS model is a density distribution based potential for the spiral arms. In this case, unlike the very simple
spiral arms mathematical approximation represented by the TWA model, PERLAS is formed, like
bricks in a building, by inhomogeneous oblate spheroids, simulating beads on a necklace (from
there the acronym). This model presents more abrupt gravitational potential and forces 
(see figures 7 and 8 in \citealt{Antoja11}). We run two PERLAS models. Model PER1 has a set of
 parameters selected so that the spiral arms are comparable to TWA2 in terms of pitch angle ($8 \deg$) and pattern speed 
(35 kms$^{-1}$kpc$^{-1}$). However, they produce a smaller force. This can be quantified with the parameter $Q_t (R)$ which 
measures the maximum azimuthal force in a given radius scaled to the axisymmetric force at that radius. While at a characteristic 
radius of $8$ kpc this parameter is 0.005 for PER1, it is 0.017 for TWA1, TWA2 and TWA3.
For model PER2 we used the same initial parameters as in PER1 but a pitch angle of 15.5 deg. instead of 8 deg. and the pattern speed is 
fixed to 20 kms$^{-1}$kpc$^{-1}$. Because of the larger pitch angle, the torque produced by PER2 is higher than for PER1 and it has a 
parameter $Q_t$ at 8 kpc of 0.020, which makes it more similar in terms of force to the previous TWA models.

\begin{figure*}
\includegraphics[scale=0.3]{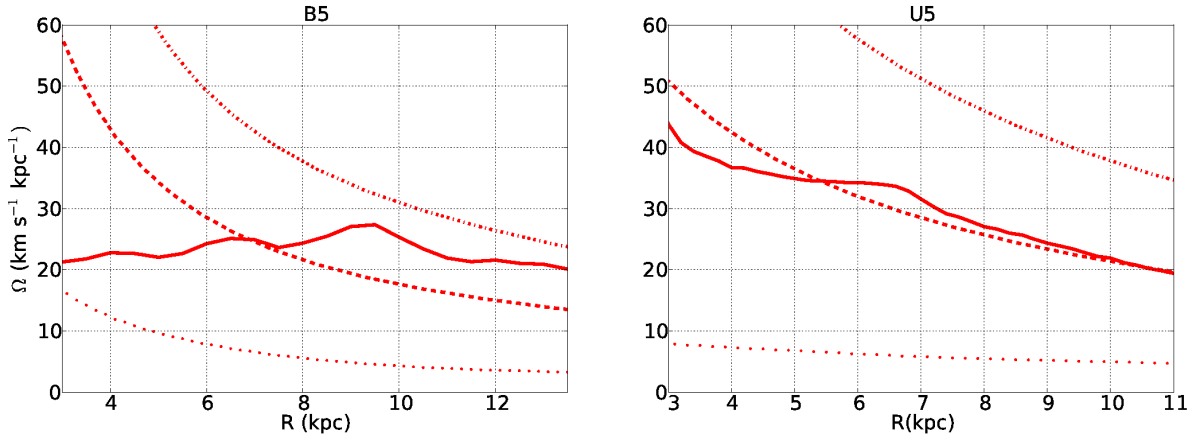}
\caption{Disk (red dashed) and bar and spiral pattern (red solid line) rotation frequency as a function of radius for the N-body model B5 (left) and U5 (right). 
The ILR and OLR curves are shown as red dotted and red dot-dashed, respectively.}
\label{fig:fig0}
\end{figure*}

\subsubsection*{Barred potentials}\label{subsec:barred}

We selected barred potentials that generates a spiral arm structure as a response.
 Several tests have been performed using Ferrers and Quadrupole bars.
As initial conditions we used a 3D axisymmetric Miyamoto-Nagai density 
profile with a radial velocity dispersion at the Suns' position of $\sigma_u$(R$_{\odot}$)=30 kms$^{-1}$ and a scale height value of 
$h_z$=300pc. According to \citet{Binney08} and \citet{Robin86}, these values would correspond to a hot population of Red Clump K
 giants. We refer to this set of initial conditions as ICMN30.

The characteristics of the bar potential presented here (FBar) can be found in \citet{Romerogomez11}. 
As this model has been developed to be compared with the Milky Way its main parameters are
 fixed within observational ranges for the Milky Way bar \citep[see][]{Romerogomez11}. The model is a superposition of two bars
 in order to obtain a boxy/bulge type of bar. For the COBE/DIRBE bulge we set the semi-major axis to $a=3.13$ kpc and the axes ratios 
to $b/a=0.4$ and $c/a=0.29$. The mass is $M_{bul}=6.3\times\,10^{9}\,M_{\odot}$. The length of the other bar, that is a Long bar,
 is set to $a=4.5$ kpc and the axes ratios to $b/a=0.15$ and $c/a=0.026$. The mass of the bar
 is fixed to $M_b=3.7\times\,10^{9}\,M_{\odot}$. This bar is introduced adiabatically and its mass is subtracted from the one of the 
Allen \& Santillan bulge \citep{Romerogomez11}. The total mass is, therefore, $M_b=10^{10}\,M_{\odot}$. In this case, the bar rotates
 at a constant pattern speed of 50 kms$^{-1}$kpc$^{-1}$.

\begin{table}
    \tabcolsep 4.pt
\begin{tabular}{lccccccc}
\hline
\hline
Model & i & Initial & N & $\Omega_b$ & $R_{CR}$ & $t_{int}$ \\
 & (deg) & Conditions & ($10^6$) & (kms$^{-1}$kpc$^{-1}$)& (kpc) &  \\
\hline
TWA1 & 8 & ICMN20 & 5 & 20&10.2 & 5 rot.\\
TWA2 & 8 & ICMN20 & 5 & 35&6.2 & 5 rot.\\
TWA3 & 8 & ICMN20 & 5 & 50&4.04 & 5 rot.\\
PER1 & 8 & ICMN20 & 4.3 & 35&6.2 & 5 rot.\\
PER2 & 15.5 & IC2 & 4.8 & 20&10.2 & 5 rot.\\
FBar & - & ICMN30 & 80 & 50&4.04 & 18 rot.\\
B5 & - & - & 5 & 22 &7.7 & 1.2 Gyr\\
U5 & - & - & 5 & -&- & 1 Gyr\\
\hline
\end{tabular}
\centering
\caption{Parameters of test particle and N-body simulations. TWA, PER and FBar are test particle models with imposed cosine 
spiral arms, PERLAS spiral arms and Ferrers' bar potential, respectively (see Sect.~\ref{subsec:test}). B5 and U5 are two snapshots from 
different N-body simulations (see Sect.~\ref{subsec:nbody}). The i values refer to the pitch angle of the imposed spiral structure, $\Omega_b$
 the pattern speed of the perturbation, $R_{CR}$ its CR and $t_{int}$ the integration time of the simulation. }
\label{tab:simulations}
\end{table}

      \subsection{Collisionless N-body models}\label{subsec:nbody}

The N-body simulations we use are the ones presented in \citet{RocaFabrega2013} as B5 and U5 models. All of them include a
 live disk and live halo but not a gas component \citep{Valenzuela2003}. B5 model has an effective number of particles of nearly 400 million, 5 of them in the disk. We built the
model ensuring the formation of a strong bar and a bisymmetric spiral associated to it. U5 model is similar to B5 but has a heavier disk and a smaller 
halo, which inhibits the bar formation. As a consequence, in the U5 simulation
 a multi-armed structure dominates, and resembles a late type galaxy with transient high m spiral waves. Simulations like 
the ones used here have been long tested to avoid numerical effects \citep[see][]{Valenzuela2003,AvilaReese05,Klypin2009}.

   Fig.~\ref{fig:fig0} shows the rotation frequency of the disk particles and the spiral arms dominant mode (red dashed
 and red solid line, respectively), and the ILR and OLR curves (red dotted and red dot-dashed, respectively). The rotation 
frequency has been computed using the method described in \citet{RocaFabrega2013}. Basically we find the spiral density structures
using Fourier analysis and later on we compute the rotation frequency from a finite differentiation of three consecutive snapshots of
the simulation.
In the B5 model a strong bar is present up to 7.7 kpc that is where the spiral arm structure begins. Fig.~\ref{fig:fig0} left
 panel shows how the bar (that ends at CR$\sim$7.7 kpc) and the spiral arms rotate at the same nearly flat rotation frequency 
($\Omega=24\pm3$km~s$^{-1}$~kpc$^{-1}$). In the U5 model the dominant mode is the m=4 and as it can be seen in right panel of 
Fig.~\ref{fig:fig0} spirals nearly corotate with disk particles.

 The high temporal and spatial resolutions and the large number of disk particles makes U5 and B5 models one of the best available 
simulations to measure kinematic quantities in the entire galactic disk  with enough resolution.

\section{CR and OLR radius from vertex deviation patterns} \label{sec:lvres}

In this section we show the behaviour of l$_v$ across the galactic disk in our different models. For that we split the disk in cylindrical sectors
 (integrated for $\left| z \right|< 0.5$ kpc). We select each region to have a $\Delta$r=200 pc and a $\Delta\theta$=6 deg. Each region overlaps 100 pc 
and 3  deg
with the contiguous ones. The expressions used for the computation of l$_v$ and its error are given in Sect.~\ref{sec:lv}.

In the polar plots of this section the disk rotates from left to right. In all plots we overplot the locus
 of the bisymmetric spiral structure as a thick solid black line. In the analytical analysis, this is given directly by the 
equation of the density perturbation of the TWA. In the simulations we show the Fourier m=2 mode locus computed by applying 
a spatial Fourier analysis in radial bins \citep{RocaFabrega2013}. For the simulations we also show density contours of regions
 with density above the mean. We computed the overdensity value of each region by subtracting the mean radial density to the
 local value. 
We mark the spiral CR with a thick solid horizontal black line and 
the OLR radius with a thick dashed horizontal black line, if those are well defined. Note that for U5 model there is no CR as
the material is corotating with the spiral pattern and also that we do not plot the m=2 Fourier mode as this does not represent the
 spiral structure (in this case we have a four armed spiral instead).
The white regions in the l$_v$ plots correspond to regions where the relative error in l$_v$ is above $50\%$.

\subsection{TWA analytical approach}\label{subsec:vdf}

\begin{figure}
\centering
\includegraphics[scale=0.3]{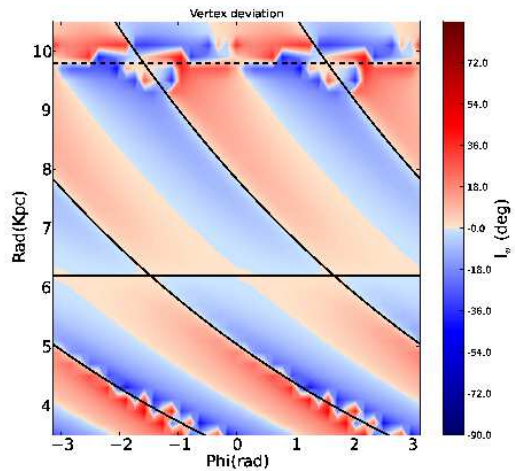}
\caption{Vertex deviation polar plots in a color scale (red for positive values, blue for negative) for the TWA analytical solution (see Sect.~\ref{subsec:analytic}).
 The solid and the dashed horizontal black lines show the position of CR and OLR 
radius, respectively. The curved black solid lines show the position of the spiral arms locus. The galaxy rotates from left to right.}
\label{fig:analylv}
\end{figure}

The results of our analytical development are presented in Fig.~\ref{fig:analylv}. This map clearly shows that 
l$_v$ follows periodic patterns related to the position of the spiral arms. In particular, it changes the sign
when moving from behind to in front of the spiral perturbation. Additionally, positions with maximum or minimum
 spiral arm potential correspond to regions with 
almost null l$_v$. This result confirms the correlation between the mass density distribution and the l$_v$, which
 was already pointed out by \citet{Vorobyov06}.

Besides, we notice here a novel result when studying the second order moments of the velocity distribution. We see that when 
crossing the spiral arms overdensity in the direction of rotation, the sign of the l$_v$ changes from positive to negative if we are inside the CR,
 but the other way around between CR and
 OLR radius, and again from positive to negative outside OLR radius.  
 Note here that \citet{Mayor70} computed l$_v$ values only at the Solar neighbourhood in a model where the Sun was placed inside CR.
 Therefore, he could not notice these patterns.
 In next sections we use this analytical result as a framework to understand the kinematics observed in our test particle and N-body
 simulations.

A question that arises from the results presented in this section is what the origin of the l$_v$ sign
 changes is. We deeply analyzed the analytical expression for the l$_v$ (see Eq.\ref{eq:lvan1}) presented in Appendix A
 to answer this question. We found that these sign changes are driven
 by the term Re$\left(i\vartheta_1\right)$D$_{\nu}^{(1)}(x)$ in the numerator. The part 
Re$\left(i\vartheta_1\right)$ drives the change that occurs when
 crossing the density peak and it corresponds to the imaginary part of the spiral arm potential, that is shifted $\pi$/2 from the spiral arm density.
The term D$_{\nu}^{(1)}(x)$, which is a function of $\nu$=m($\Omega_p$-$\Omega$)/$\kappa$,
 drives the change at CR and is related to te fact that the rotation frequencies of stars are larger or smaller than the
patterns' rotation. This is a quantitative explanation, but a qualitative physical origin of these sign changes remains unclear.

 \subsection{Results from test particle models}
 
Here we discuss the patterns of l$_v$ we obtained for all test particle models presented in Sect.~\ref{subsec:test} and its connection 
with the analytic results we show in Sect.~\ref{subsec:analytic}.

 \subsubsection{TWA spiral arms potential}

\begin{figure*}
\centering
\includegraphics[scale=0.22]{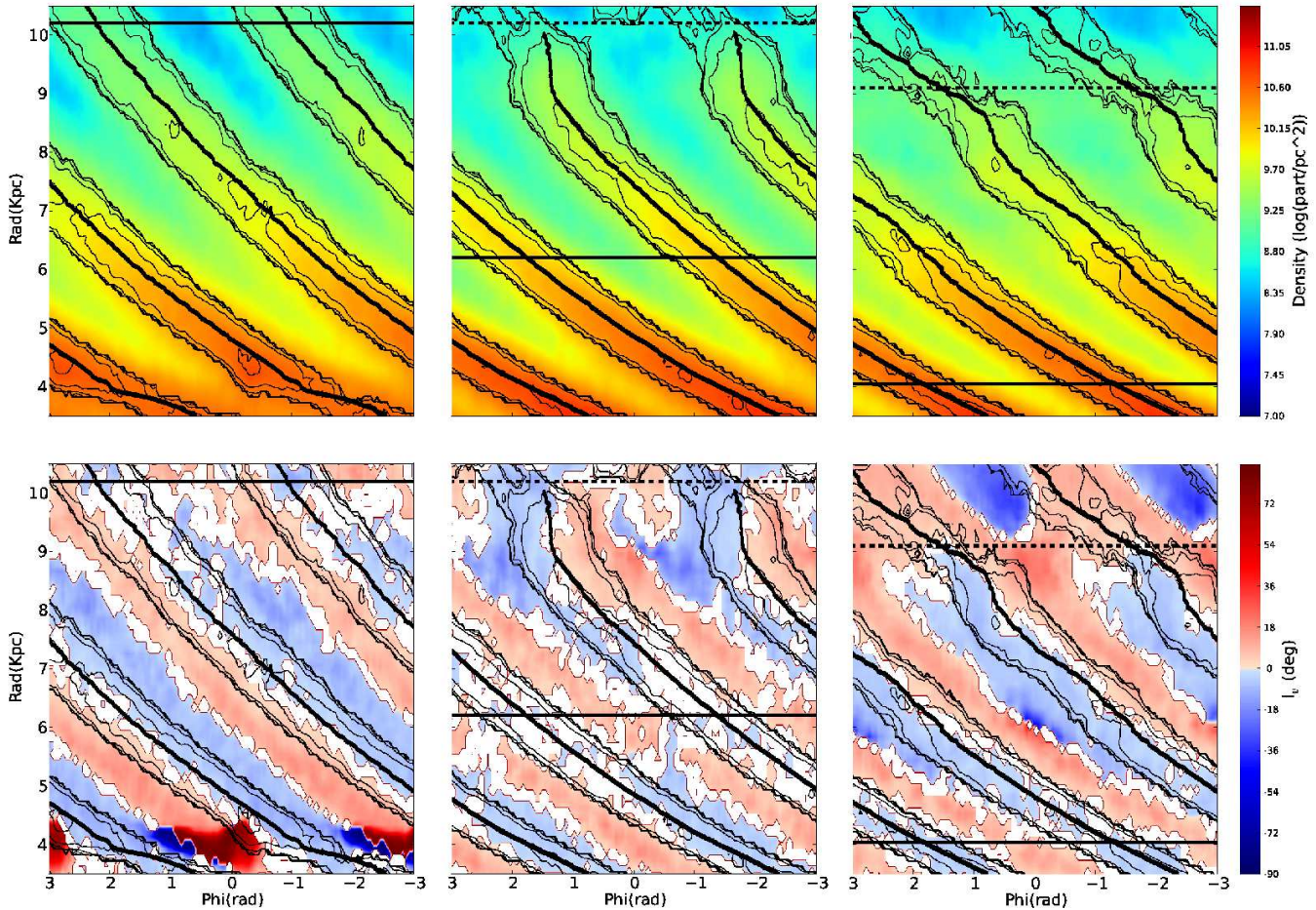}
\caption{Density (top) and l$_v$ (bottom) polar plots for test particle models TWA1 (left), TWA2 (middle) and TWA3 (right) 
from Table~\ref{tab:simulations}. The thin black lines show density contours of regions with density above the mean. The thick and dashed horizontal black lines 
show the position of CR and OLR radius, respectively. The thick black lines show the position of the Fourier m=2 mode locus.
 White regions in the bottom panels corresponds to regions where the l$_v$ relative error is above 50\%.}
\label{fig:fig1}
\end{figure*}

Fig.~\ref{fig:fig1} shows the density distribution (top) and the l$_v$ values (bottom) in polar coordinates across the whole disk
 for models TWA1 (left), TWA2 (middle) and TWA3 (right) that differ only by their spiral pattern rotation frequency. When we compare these plots with 
Fig.\ref{fig:analylv} we see that the l$_v$ structures are not so sharp. We also see that due to both, the Poisson noise and that the velocity
 ellipsoid is so rounded, the l$_v$ has a large uncertainty in some regions. As explained before, these regions with a high error in the
l$_v$ appear in white.

Here, we see the same behaviour of l$_v$ as seen in the analytical expressions of previous section. This is consistent and
expected because the underlying spiral arm potential model is the TWA in both cases. However, here we did not impose a certain 
distribution function but compute the real orbits of particles in this potential. In particular, we can clearly see this for TWA2 
which is a test particle simulation 
with similar initial conditions and parameters as the analytical approach potential. For the other two cases, TWA1 and TWA3, where most of the disk is either
 inside or outside CR we observe that the behaviour is the same as inside or outside CR, respectively, in TWA2. Note also the reverse of l$_v$ sign beyond OLR for TWA3.

\begin{figure*}
\centering
\includegraphics[scale=0.19]{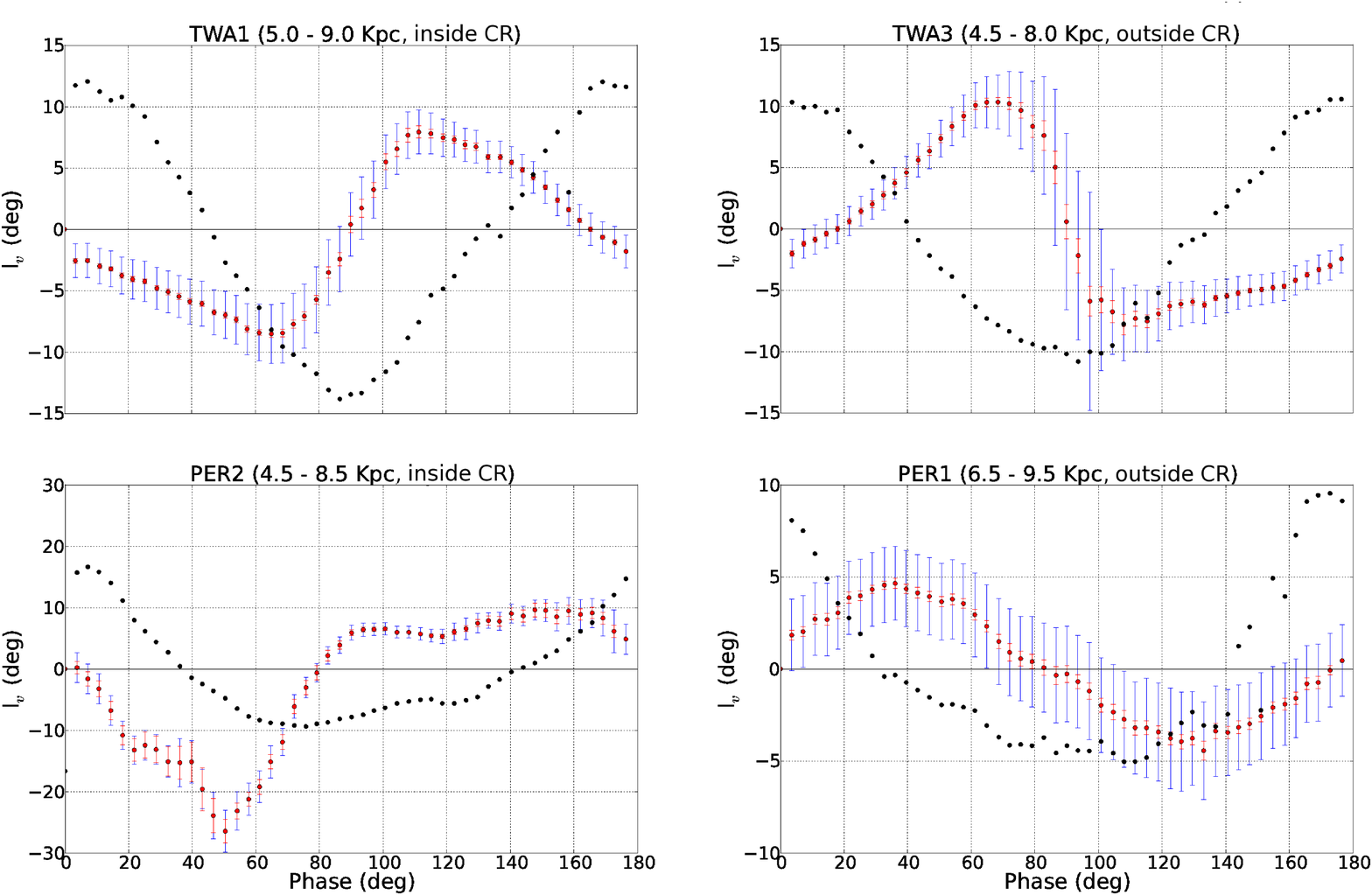}
\caption{Vertex deviation radial mean values with their errors (blue/red error bars) and spiral arm mean overdensity with respect to the mean 
disk density (black points), as a function of
angular distance to the spiral. The phase has been adjusted so that the spiral arm density peak is located at a phase of 0$^o$ while 
the minimum is at 90$^o$). Blue error bars correspond to the root mean square of errors computed using Eq.~\ref{eq:lverr}, and the red error bars
 to the error of the mean. The overdensity values have been normalized to fit in the l$_v$ plot for a better comparison of the curves.
Left: Values in regions inside CR (TWA1, top, PER2, bottom). 
Right: Regions outside CR (TWA3, top, PER1, bottom).}
\label{fig:fig6}
\end{figure*}

To see all these results in more detail we plot in Fig.~\ref{fig:fig6} (top panels)
the $l_v$ and the overdensity (black) values, as a function of angular distance to the spiral arm overdensity peak, for the models TWA1 (left) and TWA3 (right).
This distance is taken as positive in the sense of rotation. Two error bars are overplotted to the $l_v$ points. The blue ones show the root mean square of the errors obtained from Eq.~\ref{eq:lverr},
 so they reflect the Poisson noise (low number of particles) and the uncertainty when the velocity ellipsoid is almost circular.
 The red error bar is simply the error of the mean, that is the standard deviation divided by the square root of the number of regions.
 It accounts for the spread on l$_v$ at a given angular distance in the radial interval considered.
The l$_v$ for TWA1 follows an oscillation from negative values in front of the
spiral arm (for phases 0 to $\pi/2$), through null l$_v$ in the interarm region ($\pi/2$), to positive values when
approaching the next spiral arm from behind (for phases $\pi/2$ to $\pi$). If we compare the l$_v$ oscillation with the one of the overdensity 
we can conclude that the former is shifted about $\pi$/4 towards to smaller angles. An opposite shift is observed for TWA3. Note also that a 
small shift is present at 0 phase: l$_v$ is not exactly zero. This shift can be a consequence of the difficulty of finding the
density peak properly as it is not a simple sharp peak.

We also point out here that TWA1 and TWA3 models show a clear antisymmetry with respect to interarm region (angular distance from the 
spiral equal to $\pi$/2). This is a consequence of the symmetries of the potential, similar to what happens in a barred model where
 there is four-fold symmetry \citep{Fux01}. In this case, the symmetries are related precisely to the phase with respect to the 
spiral arm and that is why it appears in this maps.

After our exploration of parameters detailed in Sect.~\ref{subsec:test} we conclude that the behaviour of the l$_v$
 presented here is independent of the parameters of the TWA potential and of the initial conditions. We also observe that when imposing
 lower velocity
 dispersions in the initial conditions the $l_v$ signatures have a better definition than when we use higher velocity dispersions.

  \subsubsection{PERLAS spiral arms potential}\label{subsec:perl}

\begin{figure*}
\centering
\includegraphics[scale=0.27]{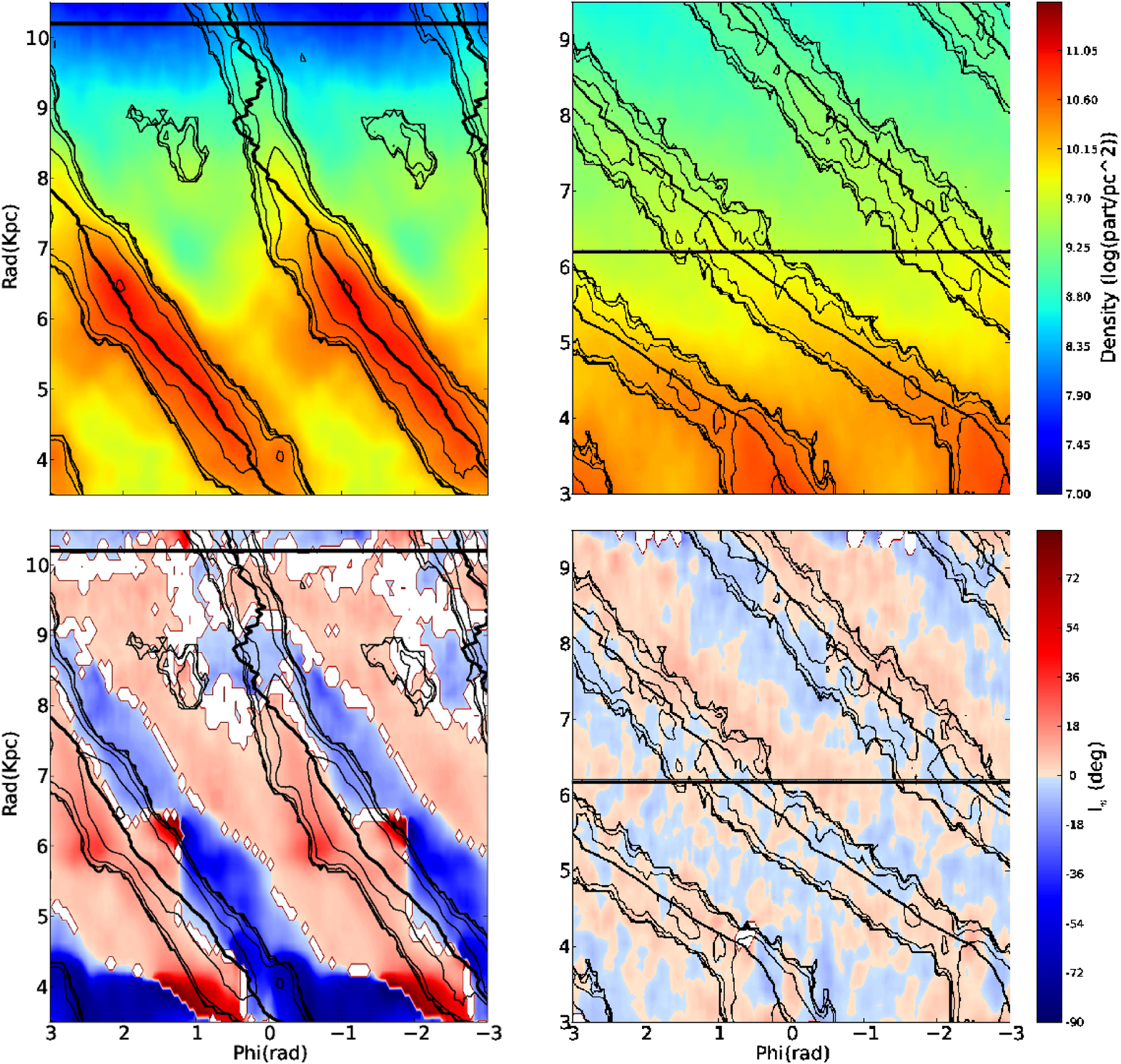}
\caption{Density (top) and l$_v$ (bottom) polar plots for test particle spiral arm models PER2 (left) and PER1 (right) from Table~\ref{tab:simulations}. See Fig.~\ref{fig:fig1} for more details on the lines.}
\label{fig:fig2}
\end{figure*}

In Fig.~\ref{fig:fig2} we show the density (top) and the l$_v$ (bottom) in polar plots of the test particle models
 where we imposed the PERLAS spiral arm potential PER2 (left) and PER1 (right). 
 This potential is more complex than the cosine expression for the force of the TWA and, as a consequence,
 the density structures appearing in these models are more complex.
 One can see, for instance, a bar-like structure in the inner radius for PER1 or the two overdensities outside
 the spiral arms at radius between 8 and 9 kpc for PER2.

The l$_v$ analysis reveals that, as in the TWA models, there is a clear relation between this 
parameter and the density structures.
In PER1, outside CR we observe the same l$_v$ pattern as in TWA analytical and test particle 
models in the same region: positive sign in front of spirals and negative sign behind them. Inside CR, however, 
we do not see a clear behaviour due to the fact that the l$_v$ values are small and that in these regions there are many 
density substructures. We point out that the presence of these density substructures (apart from the main imposed spiral arms)
 explains the higher uncertainty that exists around the CR.

For PER2, which is inside CR in the shown range of radius, in general we observe the same l$_v$ behaviour as
 in the analytical solution and in the TWA models inside CR. Note however, that between 8 and 9 kpc, where additional
 overdensities showed up, the l$_v$ appears distorted. Note also that the magnitude of the l$_v$ 
is much smaller for PER1 than for PER2, as correspond to its smaller force amplitude (see Sect.~\ref{subsec:test}).

The results for PER models become more clear in Fig.~\ref{fig:fig6}, bottom panels, where we plot l$_v$ values as a function of angular
distance to the spiral arm overdensity peak. The general behaviour of the oscillation for PER1 is similar to TWA3 (outside CR), 
although the magnitude of the l$_v$ is smaller as correspond to a smaller Q$_t$(R) (Sect.~\ref{subsec:perl}). For PER2, the oscillation
resembles that of the TWA1 (inside RC) in a first approach. However, the detailed shapes of the curves of models TWA and 
PER are slightly different. Again this must be due to the differences in the force fields, in particular in the shape of the forces as a 
function of the position in the disk mentioned in Sect.~\ref{subsec:test}. Note also that a comparison between these two different
 models was done in \citet{Antoja11} who concluded that, even when models with the same spiral locus, amplitude of the force and 
pattern speed were used, the obtained velocity field could be significantly different in some parts due to the difference in the 
force field.
As an example, the antisymmetry in the $l_v$ distribution observed in TWA models is clearly broken in the PER2 case. As seen in 
\citet{Pichardo03}, in the PERLAS model forces are not symmetric with respect to the spiral arm locus, that is the ones in front 
of the spiral are different from the ones behind.

\subsubsection{Ferrers bar potential}

\begin{figure}
\centering
\includegraphics[scale=0.19]{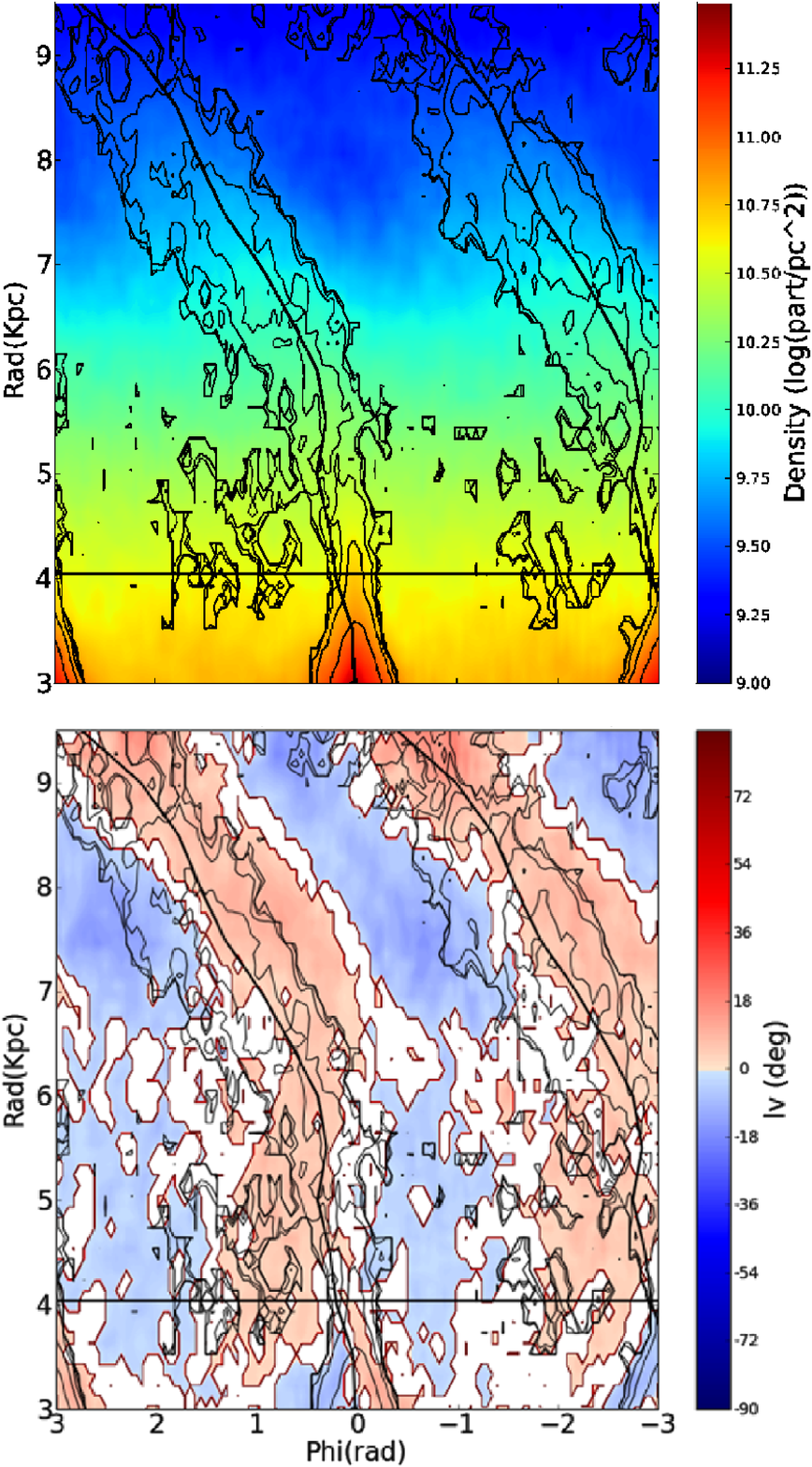}
\caption{Density (top) and l$_v$ (bottom) polar plots for model FBar from Table~\ref{tab:simulations}. See Fig.~\ref{fig:fig1} for more details on the lines.}
\label{fig:fig3}
\end{figure}

Here we analyse spiral arm structures that are produced as a response of a 3D bar potential. In Fig.~\ref{fig:fig3} we present the density (top)
 and l$_v$ (bottom) polar maps for a test particle simulation where we imposed a Ferrers bar potential (FBar). The spiral arms generated
 in this simulation can be observed in the top panel as the diagonal structures. The vertical straight structure between 3 and 4 kpc is the bar
 whereas a ring-like structure is formed between 4 and 6 kpc. As a consequence of its nature, the spirals formed in these models
 have low amplitude and are placed outside CR.

\begin{figure}
\centering
\includegraphics[scale=0.21]{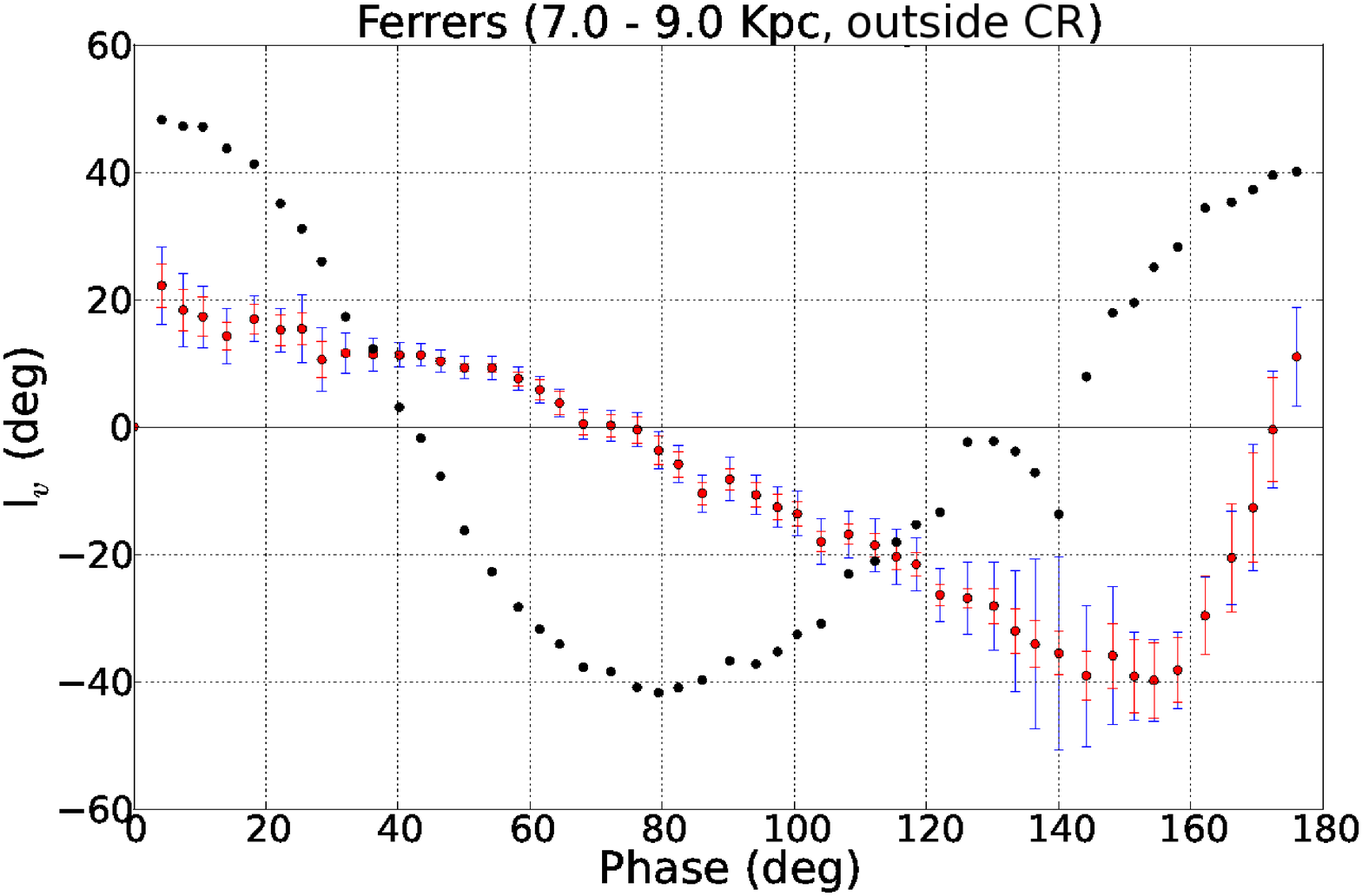}
\caption{Vertex deviation radial mean values with their errors (blue/red points) and spiral arm overdensity (black points) as function of distance to the spiral arms density peak,
 for the Ferrers bar model i.e. outside CR. For more details see Fig.~\ref{fig:fig6}.}
\label{fig:fig7}
\end{figure}

As can be seen in Fig.~\ref{fig:fig3} (bottom) the ring region presents a complicated l$_v$ pattern. Out of it, that is at radius
 $>$ 7.0 kpc the spirals created as a response of the bar are faint (low amplitude) but well defined. In Fig.~\ref{fig:fig7} we present the oscillating
 pattern of the l$_v$ and the overdensity induced by these spirals in the radial interval 7 $<$ R $<$ 9 kpc. We observe positive vertex
 deviation structures in front of the spirals and negative deviation behind them, with almost null values near the locus (0,$\pi$) and in the
 interarm region ($\sim \pi/2$). This pattern is clearly shifted $\sim \pi/4$ to the density pattern. This behaviour is in agreement with the
 trend observed in TWA3 and PER1 models (see Fig.~\ref{fig:fig6} right). 

\subsection{Results from N-body models}\label{sec:lvnbmodels}

In these models we have a more complex scenario as the 
gravitational potential has not been imposed but it is generated by the system particles themselves. As a consequence,
 of this self-generation
there are several density structures interacting with each other through gravity. Moreover, as discussed in \citet{RocaFabrega2013}, 
these are time dependent structures (e.g. 
spiral arms are transient), which introduces an additional parameter when analysing the  l$_v$ maps.

\begin{figure*}
\centering
\includegraphics[scale=0.25]{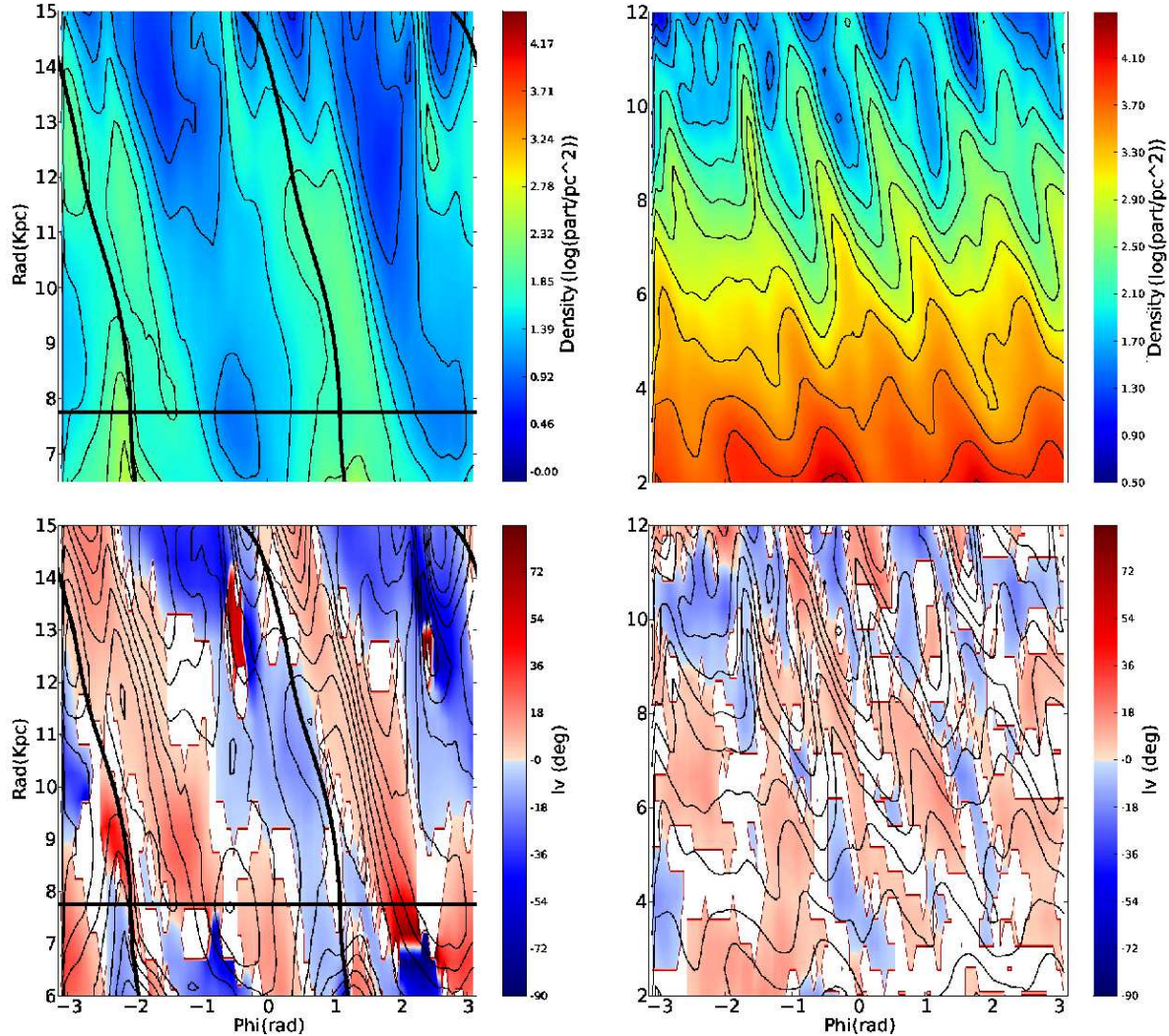}
\caption{Density (top) and l$_v$ (bottom) polar plots for N-body model B5 (left) and U5 (right) at 1.06 Gyrs of evolution. See Fig.~\ref{fig:fig1} for more details on the lines. }
\label{fig:fig4}
\end{figure*}

\begin{figure}
\centering
\includegraphics[scale=0.19]{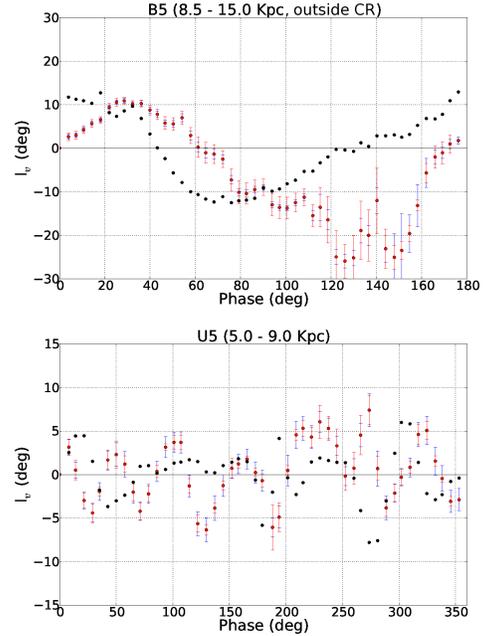}
\caption{Vertex deviation radial mean values with their errors (blue/red points) and spiral arm overdensity (black points). For more details see Fig.~\ref{fig:fig6}. Top: B5 model, i.e. outside CR; values are plotted as a function of the angular distance
to the spiral arm density peak. Bottom: U5 model, i.e. corotant structure; values are plotted for all the [0,360] angular distance range, with origin at an arbitrary angle.}
\label{fig:fig8}
\end{figure}

As mentioned in Sect.~\ref{subsec:nbody} the B5 N-body simulation develops a strong bar and a
dominant bisymmetric spiral arms which rotates roughly as a rigid body (see Fig. \ref{fig:fig0}). The spirals in this simulation are placed outside 
CR. In this model we clearly observe that, in agreement with the behaviour found in previous sections, 
positive values are found in front of the spiral and negative values are found behind (Fig.~\ref{fig:fig4}, left, and Fig.~\ref{fig:fig8}, top).
 Another interesting feature of this model is the
presence of a slow rotating m=2 mode at large galactic radius (R $\sim$ 12-14 kpc) (see \citet{RocaFabrega2013}). This is seen in Fig.~\ref{fig:fig4}, left,
where there seem to be a bifurcation in the arms at outer radius or the presence of additional arms that are not in phase with 
the main ones. These new arms rotate slower than the disk with a frequency of about 8 kms$^{-1}$kpc$^{-1}$ and they produce their own signature in l$_v$: see the additional 
two red regions at $\phi\sim-0.5$ and $\phi\sim2.5$ rad. and radius of $R\sim12.5$ kpc. These arms do show the same behaviour as previous 
arms inside CR, i.e. negative sign in front of the density perturbation and positive behind.

The second model that we analyse here is the U5 simulation. This simulation develops a multiple armed system with Fourier dominant mode being m=4,
 without a bar, and it corotates with the disk particles. The amplitude in the density of the arms in U5 is much smaller 
($A_4/A_0 \sim 0.08$) than in B5 ($A_2/A_0 \sim 0.5$). We present the results for U5 model in Figs.~\ref{fig:fig4} right, and \ref{fig:fig8} bottom. Although the amplitude
 of the l$_v$ pattern is small in this case, the small error bars allow us to provide indications that a periodicity is also present.
 However, in this case, there is no clear relation between the l$_v$ structures and density pattern. This behaviour is completely different from the B5 presented before.

Finally we make a first attempt to analyse the evolution of the l$_v$ behaviour when the density structures evolve in time 
in our N-body simulations\footnote{See movies in http://www.am.ub.edu/$\sim$sroca/Nbody/movies/.}. We find that each density structure generates its own l$_v$ pattern.  
For model B5 we see that the conclusions presented here are valid when strong spiral arms are present. Otherwise, when complex
 density structures appear, the relation between them and the l$_v$ is not straightforward. The same stands for model U5,
 with corotating spirals.

\section{Conclusions}\label{sec:conc}

In this paper we analysed the l$_v$ in simulated galactic disks with spiral structure. We mapped the l$_v$ all
 across the disk using a TWA analytical solution, several test particle simulations with imposed spiral or bar potentials and, for
 the first time, high-resolution N-body simulations. Our main outcomes are:
\begin{itemize}
 \item We confirm that the l$_v$ is clearly related to the density structure when the spiral arms are non-corotating. 
 \item In all cases with non-corotating spiral arms, the sign of the l$_v$ changes when crossing the density peak of the spiral structure
 and in the interarm region. When crossing the density peak this change is from negative to positive between CR and OLR radius and the other way around inside CR and outside OLR.
 \item When the spiral arms are corotating, there is no clear correlation between the l$_v$ and the overdensity.
\end{itemize}

Using test particle simulations we have exhaustively checked that these conclusions hold both for spiral arm potential (TWA, PERLAS) and spiral arms that are the response to an imposed bar potential
(Ferrers, quadrupole).
 Furthermore, they are independent of the initial parameters, thus on the changes of the pitch angle, the amplitude of the spirals,
 the velocity dispersion of the population or the total integration time. All these cases consist of a rigid rotating pattern with a well defined CR
 and ILR and OLR resonances. Moreover, for the first time we show here that our selfconsistent high-resolution N-body simulation with a rigid rotating 
bisymmetric and well defined two armed spiral shows a l$_v$ behaviour with the same main trends as observed in the test particle simulations.

From these models, we conclude that the changes on the sign in $l_v$ when crossing the overdensities and underdensities of the
 spiral arms give us robust and useful information about the position of the main resonant radii, that is CR and OLR. Measuring the sign
 of the l$_v$ in front or behind the spiral structure in a certain radius and azimuth would indicate whether that region 
of the galaxy is inside or outside of CR. A reverse of the sign behaviour at a certain radius would mark the CR, and in turn, give
 an estimation of the pattern speed of the spiral arms. As the position of spiral arms inside or outside CR is related to their nature
 (e.g. manifold spiral arms are generated always outside CR),
 the mapping of the l$_v$  would also trace the nature of the spiral arms in a galaxy. 

Second order differences in the shape and magnitude of l$_v$ patterns are observed when comparing all the models analyzed here. These
irregularities may be due to the intrinsic differences among their corresponding force fields. This is a matter that
 deserves further investigation but it is out of the scope of the present study.

One may wonder why models with different nature such as the TWA (low amplitude approach), PERLAS (self-gravitating 
imposed potential), response spirals in test particles barred models or N-body simulations with a well defined spiral pattern (transient structures, self-consistent model)
 show the same general trends for the l$_v$. The explanation could come from the fact that the l$_v$ is a first order effect of the velocity field, so its behaviour is successfully reproduced in all our models.
 We may require N-body simulations with a larger number of particles in order to populate the velocity distribution tails
and disentangle the differences between models through higher order momenta.

Other aspects should be addressed in a forthcoming paper in order to use our proposal as a new method to find CR and OLR.
First, an evident caveat is that the patterns in the sign of the l$_v$ are degenerate for inside CR and outside OLR. 
In this case, one would need other kinematic signatures to differentiate between those two cases.
 Second, when spiral arms coexist with other density structures (i.e. rings, floculent structures, ...), the behaviour of 
the l$_v$ can be much more complex. In these cases it will be more difficult to apply the method. Finally as it is known that the mean
velocities can also be good tracers of the CR \citep[see][]{Binney08}, it has to be studied in which cases one method is better than the other
and how they can be used complementary.

In particular for our own Galaxy, where the spiral structure is one of the main debated features in Galactic studies, it is still pending if
the Sun is inside or outside spirals' CR. Measurements of the l$_v$ at several kpc from the Solar 
neighbourhood are expected with forthcoming large surveys like Gaia \citep[ESA,][]{Perryman01}, LSST or the APO Galactic Evolution
 Experiment \citep[APOGEE-SDSS][]{Majewski10}.
The work presented here is offering new strategies to exploit this data. 

For external galaxies, as far as we know, there are no measurements of the l$_v$. In fact,
 in studies of the kinematics of external galaxies it is generally assumed that there is alignment of one of the axis of the 
stellar velocity ellipsoid with the azimuthal coordinate (i.e. there are no l$_v$) in order to derive 
properties such as the ratio between radial and azimuthal velocity dispersions from LOS velocity measurements \citep{Gerssen12,
Westfall11}. Our simulations would allow one to quantify how this assumption could bias their final results.
 Second, they can be used to establish the level of detection of l$_v$ signals in external galaxies and study which are
 the observational requirements and perspectives for detecting them with current and future instruments like ELT in its spectroscopic phase
or possible nano-arcsecond post-Gaia missions.\\

    

We thank A. Klypin and A. Kravtsov for providing us the numerical codes and L. M. Widrow for providing the code to 
generate the initial conditions.
We thank HPCC project and T. Quinn for the implementation of TIPSY package. 
This work was supported by the MINECO (Spanish Ministry of Economy) - FEDER through grant AYA2009-14648-C02-01, AYA2010-12176-E, 
AYA2012-39551-C02-01 and CONSOLIDER CSD2007-00050. SR was supported
 by the MECD PhD grant 2009FPU AP-2009-1636. SR also acknowledges the GREAT-ESF for the Short Visit Grant with ref.num 5121. TA acknowledges funding support from the European
Research Council under ERC-StG grant 
GALACTICA-240271. Simulations were carried out using HTCondor in the Kapteyn Astronomical Institute, Pakal, Abassi2 and Atocatl 
at IA-UNAM and Pirineus at CESCA.


\bibliographystyle{mn2e}

\def\apj{ApJ}
\def\apjl{ApJ}
\def\aj{AJ}
\def\mnras{MNRAS}
\def\aa{A\&A}
\def\nat{nat}
\def\araa{ARA\&A}
\def\aap{A\&A}


\bibliography{biblio}
\IfFileExists{\jobname.bbl}{}
{\typeout{}
\typeout{****************************************************}
\typeout{****************************************************}
\typeout{** Please run "bibtex \jobname" to optain}
\typeout{** the bibliography and then re-run LaTeX}
\typeout{** twice to fix the references!}
\typeout{****************************************************}
\typeout{****************************************************}
\typeout{}
}

\bsp

\newpage

 \onecolumn

\appendix

\section{Analytical expressions for the vertex deviation using the TWA approach}\label{sec:appendix1}

Here we compute the analytical values for the vertex in the whole galactic disk plane (z=0) when imposing \citet{LinShu} classical spiral arms.
 We started from the classical definition (see Eq.~\ref{eq:lv1}) and  we computed up to second order moments of the velocity 
distribution function. Finally, following \citet{Lin69} we derived an expression similar to the one presented in \citet{Mayor70} (see Eq.~\ref{eq:lvan1}).

\subsection{Notation}\label{subsec:notation}
Here we detail the notation we will use in next sections.

\begin{eqnarray}
\label{eq:notation}
\Psi &=& \text{Total velocity distribution function,} \nonumber \\
\Psi_0 &=& \text{Schwarzschild velocity distribution function,} \nonumber \\
\sigma_{*}^0 &=& \text{Non-perturbed surface density [M$_{\odot}$~kpc$^{-2}$],} \nonumber \\
\sigma_* &=& \text{Total surface density [M$_{\odot}$~kpc$^{-2}$],} \nonumber \\
m_* &=& \text{Stellar mass. We assume it is the same for all stars [M$_{\odot}$],} \nonumber \\
\varpi &=& \text{Distance to the galactic center in the disk plane [kpc],} \nonumber \\
\theta &=& \text{Azimuthal angle in the disk plane [rad.],} \nonumber \\
\theta &=& \text{Time [s],} \nonumber \\
\Omega &=& \text{Angular velocity of a particle in a circular orbit in the axisymmetric averaged potential of the galactic disk [km~s$^{-1}$~kpc$^{-1}$],} \nonumber \\
\Theta &=& \text{Radial velocity [km~s$^{-1}$],} \nonumber \\
\Pi &=& \text{Azimuthal velocity [km~s$^{-1}$],} \nonumber \\
Z &=& \text{Vertical velocity [km~s$^{-1}$],} \nonumber \\
c_{\varpi}&=&\Theta \text{:  \:  Radial residual velocity [km~s$^{-1}$],} \nonumber \\
c_{\theta}&=&\Pi-\varpi\Omega \text{:  \:  Azimuthal residual velocity [km~s$^{-1}$],} \nonumber \\
c_{z}&=&Z \text{:  \:  Vertical residual velocity [km~s$^{-1}$],} \nonumber \\
V_{\varpi},V_{\theta} &=& \text{Radial and azimuthal mean systematic movements [km~s$^{-1}$],} \nonumber \\
\sigma_{\varpi} &=& \sqrt{\mu_{200}^{(0)}}\text{:  \:  Dispersion of the non-perturbed radial residual velocities [km~s$^{-1}$],} \nonumber \\
l_v &=& \text{Vertex deviation [rad.],} \nonumber \\
\kappa &=& 2\Omega\sqrt{1+\frac{\varpi}{2\Omega}\frac{d\Omega}{d\varpi}} \text{: \:   Epicyclic frequency [km~s$^{-1}$~kpc$^{-1}$],} \nonumber \\
\gamma &=& \frac{2\Omega}{\kappa} \nonumber \\
i &=& \text{Pitch angle of the spiral [rad.],} \nonumber \\
m &=& \text{Number of spiral arms,} \nonumber \\
\omega &=& \text{Rotation frequency of the spiral [km~s$^{-1}$~kpc$^{-1}$],} \nonumber \\
\Omega_p &=& Re(\omega)/m \text{Angular velocity of the spiral arm pattern [km~s$^{-1}$~kpc$^{-1}$],} \nonumber \\
R_0 &=& \text{Initial radius for the spiral perturbation [kpc],} \nonumber \\
\Phi &=& -\frac{2}{\tan i}\ln\left(\frac{\varpi}{R_0}\right) \text{: \:   Spiral arm locus,} \nonumber \\ 
K &=&\frac{d\Phi}{d\varpi} \text{:  \:  Wave number [kpc$^{-1}$],} \nonumber \\
A_{sp} &=& \text{Spiral arms potential amplitude normalization [km$^2$~s$^{-2}$~kpc$^{-1}$],} \nonumber \\
R_{\Sigma} &=& \text{Spiral arms radial scale length [kpc],} \nonumber \\
A &=& -A_{sp}\varpi\exp^{-\varpi/R_{\Sigma}}\text{: \:  Amplitude of the spiral arms potential [km$^2$~s$^{-2}$],} \nonumber \\
\vartheta_1 &=& A \exp^{i\left(\omega t-2\theta+\Phi(\varpi)\right)} \text{:  \:  Spiral arm potential [km$^2$~s$^{-2}$],} \nonumber \\
x &=& K^2\frac{\sigma_{\varpi}^2}{\kappa^2} \text{: \: Toomre number,} \nonumber \\
\nu &=& \frac{m\left(\Omega_p-\Omega\right)}{\kappa} \text{;}\hspace{0.5cm} 
V_1\hspace{0.2cm}=\hspace{0.2cm}\frac{\left(2\Omega\right)\left(\varpi\Omega\right)}{\kappa} \text{;}\hspace{0.5cm} 
a\hspace{0.2cm}=\hspace{0.2cm}\frac{\left(K\varpi\right)\left(2\Omega^2\right)}{\kappa^2} \text{;}\hspace{0.5cm}
\mu_0\hspace{0.2cm}=\hspace{0.2cm}\frac{V_1^2}{\sigma_{\varpi}^2}\nonumber \\
\xi&=&\frac{c_{\varpi}}{V_1}\text{;}\hspace{0.5cm} \eta\hspace{0.2cm}=\hspace{0.2cm}\frac{c_{\theta}}{\varpi\Omega}  \text{: \:  Dimensionless velocities referred to local values,} \nonumber \\
\left<f\right> &=& \frac{m_*}{\sigma_{*}^0}\int\int\int{f\Psi_0dc_{\varpi}dc_{\theta}dc_z} \text{: \:  Weighed average with respect to $\Psi_0$} \nonumber
\end{eqnarray}

\subsection{Velocity distribution function}
To compute the moments we used the velocity distribution function first presented in \citet{Lin69}. This function (Eq.~\ref{eq:veldisfunc}) is a 
direct summation of a classical Schwarzschild distribution ($\Psi_o$) and a small perturbation due to the presence of a tightly wound spiral ($\Psi_1$). 
For details on the derivation of $\Psi_1$ see Appendix A in \citet{Lin69}.

\begin{eqnarray}
\label{eq:veldisfunc}
\Psi &=& \Psi_0 + \Psi_1 \nonumber \\
\Psi_0 &=& P_0(\varpi)\exp^{-\frac{\mu_0}{2}\left(\xi^2+\eta^2\right)} \nonumber \\
\Psi_1 &=& \frac{-\vartheta_1}{\sigma_{\varpi}^2}\cdot\Psi_0\cdot\left(1-q\right) \\
\text{ where:} \nonumber \\
q &=& \frac{\nu\pi}{\sin\left(\nu\pi\right)}\cdot\frac{1}{2\pi}\int^{\pi}_{-\pi}{\exp^{i\left[\nu\alpha-a\cdot\xi\sin\alpha+a\cdot\eta\left(1+\cos\alpha\right)
\right]}d\alpha}  \nonumber
\end{eqnarray}

\subsection{Computation of the moments}
The equations presented in this section have been obtained imposing the perturbed velocity distribution function (Eq.~\ref{eq:veldisfunc}) to the general expression for the moments
, Eq.~\ref{eq:mom}. To get the final expressions shown here we used the relations proposed in Sect.~\ref{subsubsec:parenthesis} and also that the first order moments
 and the crossed
 second order moments of a non-perturbed Schwarzschild velocity distribution function are 0 $(\mu_{100}^{(0)}=\mu_{010}^{(0)}=\mu_{110}^{(0)}=0)$. The epicyclic approximation was used to link the second order moments of the
 non-perturbed Shchwarzschild velocity distribution function $(\mu^{(0)}_{020}=\mu^{(0)}_{200}(2\Omega/\kappa)^{-2})$. A detailed example of how we obtain the final expressions can be seen in Eq.~\ref{eq:0mom1} and \ref{eq:1mom1}.
\subsubsection{Zero order moments:}
To obtain the expressions for the zero order moments we used the Eq.~\ref{eq:parenth1}.
\begin{eqnarray}
\label{eq:0mom}
\sigma_{*}^0 &=& \mu_{000}^{(0)} = m_*\int\int\int{\Psi_0 dc_{\varpi}dc_{\theta}dc_z}
\end{eqnarray}
\begin{eqnarray}
\label{eq:0mom1}
\sigma_{*} &=& \mu_{000}~m_* = m_*\int\int\int{\Psi dc_{\varpi}dc_{\theta}dc_z} = m_*\int\int\int{\Psi_0 + \Psi_0 \frac{-\vartheta_1}{\sigma_{\varpi}^2}(1-q) dc_{\varpi}dc_{\theta}dc_z}= \nonumber \\
 &=& \sigma_{*}^0\left(1-\frac{\vartheta_1}{\sigma_{\varpi}^2}\left<(1-q)\right>\right)=\sigma_{*}^0 \left[1-\frac{\vartheta_1}{\sigma_{\varpi}^2}\left(1-\frac{\nu\pi}{\sin\left(\nu\pi\right)}\frac{1}{2\pi}\int^{\pi}_{-\pi}\cos\left(
\nu\alpha\right)\exp^{-\frac{a^2}{\mu_0}\left(1+\cos\alpha\right)}d\alpha\right)\right] 
\end{eqnarray}

\subsubsection{First order moments:}
We obtained the expressions for the first order moments using Eq.~\ref{eq:parenth2} and \ref{eq:parenth3}, after changing the residual velocities $(c_{\varpi},c_{\theta})$ for the
 dimensionless velocities referred to local values $(\xi,\eta)$. We also used that $\mu_{100}^{(0)}=\mu_{010}^{(0)}=0$.

\begin{eqnarray}
\label{eq:1mom1}
V_{\varpi} \;\:=\;\: \mu_{100} &=& \frac{m_*}{\sigma_{*}}\int\int\int c_{\varpi}\Psi dc_{\varpi}dc_{\theta}dc_z= \nonumber \\
 &=& \frac{\sigma_{*}^0}{\sigma_{*}}\left[\frac{m_*}{\sigma_{*}^0}\int{dc_z}\int\int c_{\varpi}\Psi_0 dc_{\varpi}dc_{\theta}-\frac{\vartheta_1}{\sigma_{\varpi}^2}\frac{m_*}{\sigma_{*}^0}\int{dc_z}\int\int c_{\varpi}\Psi_0\left(1-q\right)dc_{\varpi}dc_{\theta}\right]=  \\
 &=& - \frac{\sigma_{*}^0}{\sigma_{*}}\frac{\vartheta_1}{\sigma_{\varpi}^2}\frac{2\Omega\left(\varpi\Omega\right)}{\kappa}\left<\xi\left(1-q\right)\right> = 
\frac{\sigma_{*}^0}{\sigma_{*}}\frac{\vartheta_1K}{\kappa}\frac{\nu\pi}{\sin\left(\nu\pi\right)}\frac{1}{2\pi}\int^{\pi}_{-\pi}\sin\left(\nu\alpha\right)\sin\alpha\exp^{-\frac{a^2}{\mu_0}
\left(1+\cos\alpha\right)}d\alpha \nonumber
\end{eqnarray}

\begin{eqnarray}
\label{eq:1mom2}
V_{\theta} \;\:=\;\: \mu_{010} &=& \frac{m_*}{\sigma_{*}}\int\int\int c_{\theta}\Psi dc_{\varpi}dc_{\theta}dc_z=-i\frac{\sigma_{*}^0}{\sigma_{*}}\frac{\vartheta_1K}{2\Omega}\frac{\nu\pi}
{\sin\left(\nu\pi\right)}\frac{1}{2\pi}\int^{\pi}_{-\pi}\left(1+\cos\alpha\right)\cos\left(\nu\alpha\right)\exp^{-\frac{a^2}{\mu_0}\left(1+\cos\alpha\right)}d\alpha 
\end{eqnarray}

\subsubsection{Second order moments:}
Here we obtained the expressions for the second order moments using Eq.~\ref{eq:parenth4}, \ref{eq:parenth5} and \ref{eq:parenth6}, after changing the residual velocities 
$(c_{\varpi},c_{\theta})$ for the dimensionless velocities referred to local values $(\xi,\eta)$. We also used that $\mu_{110}^{(0)}=0$ and 
$\mu^{(0)}_{020}=\mu^{(0)}_{200}(2\Omega/\kappa)^{-2}$.

\begin{eqnarray}
\label{eq:2mom1}
\mu_{110} &=& \frac{m_*}{\sigma_{*}}\int\int\int\Psi c_{\theta}c_{\varpi}dc_{\varpi}dc_{\theta}dc_z=\frac{\sigma_{*}^0}{\sigma_{*}}\left(
\mu^0_{110}-\frac{\vartheta_1}{\sigma_{\varpi}^2}\frac{2\Omega}{\kappa}\left(\varpi\Omega\right)^2\left<\xi\eta(1-q)\right>\right)= \nonumber \\
 &=& \frac{\sigma_{*}^0}{\sigma_{*}} \left(\frac{\vartheta_1\kappa}{2\Omega}\frac{a^2}{\mu_0}\frac{\nu\pi}
{\sin\left(\nu\pi\right)}\frac{i}{2\pi}\int^{\pi}_{-\pi}\sin\left(\nu\alpha\right)\sin\alpha\left(1+\cos\alpha\right)\exp^{-\frac{a^2}{\mu_0}\left(1+\cos\alpha\right)}d\alpha\right)
\end{eqnarray}
\begin{eqnarray}
\label{eq:2mom2}
\mu_{200} &=& \frac{m_*}{\sigma_{*}}\int\int\int\Psi c_{\varpi}^2dc_{\varpi}dc_{\theta}dc_z=\frac{\sigma_{*}^0}{\sigma_{*}}\left(\mu^0_{200}-\frac{\vartheta_1}
{\sigma_{\varpi}^2}\left(\frac{2\Omega}{\kappa}\right)^2\left(\varpi\Omega\right)^2\left<\xi^2(1-q)\right>\right)= \nonumber \\
 &=&\frac{\sigma_{*}^0}{\sigma_{*}}\left(\frac{\kappa}{2\Omega}\right)^2\left(\sigma_{\varpi}^2-\vartheta_1\left[1-\frac{\nu\pi}
{\sin\left(\nu\pi\right)}\frac{1}{2\pi}\int^{\pi}_{-\pi}\left(1-\frac{a^2}{\mu_0}-2\frac{a^2}{\mu_0}\cos\alpha-\frac{a^2}{\mu_0}\cos^2\alpha\right)
\cos\left(\nu\alpha\right)\exp^{-\frac{a^2}{\mu_0}\left(1+\cos\alpha\right)}d\alpha\right]\right)
\end{eqnarray}
\begin{eqnarray}
\label{eq:2mom3}
\mu_{020} &=& \frac{m_*}{\sigma_{*}}\int\int\int\Psi c_{\theta}^2dc_{\varpi}dc_{\theta}dc_z=\frac{\sigma_{*}^0}{\sigma_{*}}\left(
\mu^0_{020}-\frac{\vartheta_1}{\sigma_{\varpi}^2}\left(\varpi\Omega\right)^2\left<\eta^2(1-q)\right>\right) \nonumber \\
 &=& \frac{\sigma_{*}^0}{\sigma_{*}}\left(\frac{\kappa}{2\Omega}\right)^2\left(\sigma_{\varpi}^2-\vartheta_1\left[1-\frac{\nu\pi}
{\sin\left(\nu\pi\right)}\frac{1}{2\pi}\int^{\pi}_{-\pi}\left(1-\frac{a^2}{\mu_0}-2\frac{a^2}{\mu_0}\cos\alpha-\frac{a^2}{\mu_0}\cos^2\alpha\right)
\cos\left(\nu\alpha\right)\exp^{-\frac{a^2}{\mu_0}\left(1+\cos\alpha\right)}d\alpha\right]\right)
\end{eqnarray}

\subsubsection{Centered second order moments:}\label{subsec:2cmom}
Expressions for the centered second order moments have been obtained using the equations obtained in the
 previous sections (\ref{eq:0mom} to \ref{eq:2mom3}).

\begin{eqnarray}
\label{eq:2momc1}
\tilde{\mu}_{110} &=& \frac{m_*}{\sigma_{*}}\int\int\int\Psi\left(c_{\theta}-V_{\theta}\right)\left(c_{\varpi}-V_{\varpi}\right)dc_{\varpi}dc_{\theta}dc_z= \nonumber \\
 &=& \frac{m_*}{\sigma_{*}}\left(\int{dc_z}\int\int\Psi c_{\theta}c_{\varpi}dc_{\varpi}dc_{\theta}-V_{\theta}\int{dc_z}\int\int\Psi c_{\varpi}dc_{\varpi}dc_{\theta}-
V_{\varpi}\int{dc_z}\int\int\Psi c_{\theta}dc_{\varpi}dc_{\theta}+V_{\theta}V_{\varpi}\sigma_{*}\right)=  \\
 &=& \frac{\sigma_{*}^0}{\sigma_{*}} \left(\frac{\vartheta_1\kappa}{2\Omega}\frac{a^2}{\mu_0}\frac{\nu\pi}
{\sin\left(\nu\pi\right)}\frac{i}{2\pi}\int^{\pi}_{-\pi}\sin\left(\nu\alpha\right)\sin\alpha\left(1+\cos\alpha\right)\exp^{-\frac{a^2}{\mu_0}\left(1+\cos\alpha\right)}d\alpha-
V_{\theta}V_{\varpi}\frac{\sigma_{*}}{\sigma_{*}^0}\right) \nonumber
\end{eqnarray}
\begin{eqnarray}
\label{eq:2momc2}
\tilde{\mu}_{200} &=& \frac{m_*}{\sigma_{*}}\int\int\int\Psi\left(c_{\varpi}-V_{\varpi}\right)^2dc_{\varpi}dc_{\theta}dc_z= \nonumber \\
 &=& \frac{m_*}{\sigma_{*}}\left(\int{dc_z}\int\int\Psi c_{\varpi}^2dc_{\varpi}dc_{\theta}+V_{\varpi}^2\sigma_{*}-
2V_{\varpi}\int{dc_z}\int\int\Psi c_{\varpi}dc_{\varpi}dc_{\theta}\right)=  \\
&=& \frac{\sigma_{*}^0}{\sigma_{*}}\left( \sigma_{\varpi}^2-\vartheta_1\left[1-\frac{\nu\pi}
{\sin\left(\nu\pi\right)}\frac{1}{2\pi}\int^{\pi}_{-\pi}\left(1-\frac{a^2}{\mu_0}\sin^2\alpha\right)\cos\left(\nu\alpha\right)
\exp^{-\frac{a^2}{\mu_0}\left(1+\cos\alpha\right)}d\alpha\right]-V_{\varpi}^2\frac{\sigma_{*}}{\sigma_{*}^0}\right) \nonumber
\end{eqnarray}
\begin{eqnarray}
\label{eq:2momc3}
\tilde{\mu}_{020} &=& \frac{m_*}{\sigma_{*}}\int\int\int\Psi\left(c_{\theta}-V_{\theta}\right)^2dc_{\varpi}dc_{\theta}dc_z= \nonumber \\
 &=& \frac{m_*}{\sigma_{*}}\left(\int{dc_z}\int\int\Psi c_{\theta}^2dc_{\varpi}dc_{\theta}+V_{\theta}^2\sigma_{*}-
2V_{\theta}\int{dc_z}\int\int\Psi c_{\theta}dc_{\varpi}dc_{\theta}\right)=  \\
&=& \frac{\sigma_{*}^0}{\sigma_{*}}\left(\frac{\kappa}{2\Omega}\right)^2\left(\sigma_{\varpi}^2-\vartheta_1\left[1-\frac{\nu\pi}
{\sin\left(\nu\pi\right)}\frac{1}{2\pi}\int^{\pi}_{-\pi}\left(1-\frac{a^2}{\mu_0}-2\frac{a^2}{\mu_0}\cos\alpha-\frac{a^2}{\mu_0}\cos^2\alpha\right)
\cos\left(\nu\alpha\right)\exp^{-\frac{a^2}{\mu_0}\left(1+\cos\alpha\right)}d\alpha\right]-V_{\theta}^2\frac{\sigma_{*}}{\sigma_{*}^0}\right) \nonumber
\end{eqnarray}

\subsubsection{Useful weighed averages with respect to $\Psi_0$}\label{subsubsec:parenthesis}
Next we present some useful relations that can be easily obtained using the expression for the weighed averages with respect to $\Psi_0$  presented 
in Sect.~\ref{subsec:notation}, $\left<\text{f}\right>$, and the Schwarzschild velocity distribution function ($\Psi_0$ in Eq.\ref{eq:veldisfunc}):

\begin{eqnarray}
\label{eq:parenth1}
\left<(1-q)\right> &=& 1- \frac{\nu\pi}{\sin\left(\nu\pi\right)}\frac{1}{2\pi}\int^{\pi}_{-\pi}\cos\left(\nu\alpha\right)\exp^{-\frac{a^2}{\mu_0}
\left(1+\cos\alpha\right)}d\alpha
\end{eqnarray}
\begin{eqnarray}
\label{eq:parenth2}
\left<\xi(1-q)\right> &=& -\frac{a}{\mu_0}\frac{\nu\pi}{\sin\left(\nu\pi\right)}\frac{1}{2\pi}\int^{\pi}_{-\pi}\sin\left(\nu\alpha\right)\sin\alpha 
\exp^{-\frac{a^2}{\mu_0}\left(1+\cos\alpha\right)}d\alpha
\end{eqnarray}
\begin{eqnarray}
\label{eq:parenth3}
\left<\eta(1-q)\right> &=& \frac{ia}{\mu_0}\frac{\nu\pi}{\sin\left(\nu\pi\right)}\frac{1}{2\pi}\int^{\pi}_{-\pi}\cos\left(\nu\alpha\right)\left(1+\cos\alpha\right) 
\exp^{-\frac{a^2}{\mu_0}\left(1+\cos\alpha\right)}d\alpha
\end{eqnarray}
\begin{eqnarray}
\label{eq:parenth4}
\left<\xi^2(1-q)\right> &=& \frac{\sigma_{\varpi}^2}{V_1^2}-\frac{1}{\mu_0}\frac{\nu\pi}{\sin\left(\nu\pi\right)}\frac{1}{2\pi}\int^{\pi}_{-\pi}
\cos\left(\nu\alpha\right)\left(1-\frac{a^2}{\mu_0}\sin^2\alpha\right)\exp^{-\frac{a^2}{\mu_0}\left(1+\cos\alpha\right)}d\alpha
\end{eqnarray}
\begin{eqnarray}
\label{eq:parenth5}
\left<\eta^2(1-q)\right> &=& \left(\frac{\kappa}{2\Omega}\right)^2\frac{\sigma_{\varpi}^2}{\left(\varpi\Omega\right)^2}-\frac{1}{\mu_0}\frac{\nu\pi}
{\sin\left(\nu\pi\right)}\frac{1}{2\pi}\int^{\pi}_{-\pi}\cos\left(\nu\alpha\right)\left(1-\frac{a^2}{\mu_0}-2\frac{a^2}{\mu_0}\cos\alpha-\frac{a^2}{\mu_0}\cos^2\alpha\right)
\exp^{-\frac{a^2}{\mu_0}\left(1+\cos\alpha\right)}d\alpha
\end{eqnarray}
\begin{eqnarray}
\label{eq:parenth6}
\left<\xi\eta(1-q)\right> &=& -\frac{a^2}{\mu_0^2}\frac{\nu\pi}{\sin\left(\nu\pi\right)}\frac{i}{2\pi}\int^{\pi}_{-\pi}
\sin\left(\nu\alpha\right)\sin\alpha\left(1+\cos\alpha\right)\exp^{-\frac{a^2}{\mu_0}\left(1+\cos\alpha\right)}d\alpha  
\end{eqnarray}

\subsection{Analytical expression for vertex deviation}
Using the expressions we obtained for the centered second order moments (Sect.~\ref{subsec:2cmom}), we find the analytical
 formula for the l$_v$ we present here.


\begin{eqnarray}
\label{eq:lvan1}
\frac{1}{2} \tan \left(2 \cdot l_v\right)&=& \frac{\tilde{\mu}_{110}}{\tilde{\mu}_{200}-\tilde{\mu}_{020}} = \frac{\frac{Re\left(i\vartheta_1\right)}{\gamma}D_{\nu}^{(1)}(x)-V_{\varpi}V_{\theta}\frac{\sigma_{*}}{\sigma_{*}^0}}
{\left(\sigma_{\varpi}^2-Re(\vartheta_1)\right)\left(1-\frac{1}{\gamma^2}\right)+Re(\vartheta_1)\left(D_{\nu}^{(2)}-\frac{D_{\nu}^{(3)}}{\gamma}\right)-
\left(V_{\varpi}^2-V_{\theta}^2\right)\frac{\sigma_{*}}{\sigma_{*}^0}} 
\end{eqnarray}
where:
\begin{eqnarray}
\label{eq:lvan11}D_{\nu}^{(1)}(x) &=& x\frac{\nu\pi}{\sin\left(\nu\pi\right)}\frac{1}{2\pi}\int^{\pi}_{-\pi}\sin\left(\nu\alpha\right)\sin\alpha\left(1+\cos\alpha\right)
\exp^{-x\left(1+\cos\alpha\right)}d\alpha \nonumber \\
D_{\nu}^{(2)}(x) &=& \frac{\nu\pi}{\sin\left(\nu\pi\right)}\frac{1}{2\pi}\int^{\pi}_{-\pi}\cos\left(\nu\alpha\right)\left(1-x\sin^2\alpha\right)
\exp^{-x\left(1+\cos\alpha\right)}d\alpha \nonumber \\
D_{\nu}^{(3)}(x) &=& \frac{\nu\pi}{\sin\left(\nu\pi\right)}\frac{1}{2\pi}\int^{\pi}_{-\pi}\cos\left(\nu\alpha\right)\left(1-x-2x\cos\alpha-x\cos\alpha^2
\right)\exp^{-x\left(1+\cos\alpha\right)}d\alpha \nonumber 
\end{eqnarray}

\label{lastpage}

\end{document}